\documentclass[12pt,preprint]{aastex}
\begin{document}
\title{The Southern Dwarf Hunt: Local Group Dwarf Candidates in the
Southern Sky\footnote{Based on observations made with the 
Isaac Newton Telescope, operated
on the island of La Palma by the Isaac Newton Group in the Spanish
Observatorio del Roque de los Muchachos of the Instituto de Astrof\'{\i}sica de
Canarias, and on observations made at Cerro Tololo Interamerican
Observatory (CTIO).  CTIO is operated by the Association of Universities
for Research in Astronomy, Inc. (AURA) under a cooperative agreement with
the National Science Foundation as part of the National Optical Astronomy
Observatories (NOAO).}}

\author{Alan B. Whiting\altaffilmark{2}}
\affil{Physics Department, U. S. Naval Academy}
\email{whiting@nadn.navy.mil}

\author{George K. T. Hau\altaffilmark{3}}
\affil{Facultad de F\'{\i}sica y Astronom\'{\i}a, Universidad
Cat\'{o}lica de Chile}
\email{ghau@astro.puc.cl}

\and

\author{Mike Irwin}
\affil{Institute of Astronomy, University of Cambridge}
\email{mike@ast.cam.ac.uk}

\altaffiltext{2}{Now at Cerro Tololo Interamerican Observatory, La
Serena, Chile, whiting@ctio.noao.edu}

\altaffiltext{3}{Now at the European Southern Observatory, Santiago,
Chile, ghau@eso.org}

\begin{abstract}
We present observations of 82 Local Group dwarf galaxy candidates, of
which 62 were chosen visually from ESO-SRC survey plates of the 
southern sky (32 of which were not previously catalogued)
and the rest suggested by various sources in the literature.  
Two are the Local Group galaxies Antlia and Cetus; nine 
are more distant galaxies, though still within a few megaparsecs;
45 are background galaxies; seven are planetary (or other emission) nebulae;
15 are reflection or other Galactic nebulae; two are galaxy clusters; 
one is a Galactic star cluster and one is a misidentified star.  
We conclude that there is
no large population of faint Local Group dwarf galaxies of any familiar
type awaiting discovery.
We point out the danger of relying on a single type of data to reach 
conclusions about an object.
\end{abstract}

\keywords{galaxies: dwarf---Local Group---planetary nebulae: general---
reflection nebulae---surveys}

\section{Dwarf Galaxies}

We describe here results of our search for Local Group
low surface-brightness (LSB) dwarf galaxies\footnote{These three categories
are formally separate.  There are large LSB galaxies and bright dwarfs,
and no doubt many of all sorts in the distance as well as nearby.  It
should be clear from the discussion below why in our work the three go
together.  For the sake of brevity we will generally call our targets
``dwarf galaxies'' and allow the ``LSB'' and ``nearby'' adjectives be implied.}.
LSB dwarfs form a population heavily selected against in galaxy surveys and 
catalogs (for obvious reasons), but of disproportionate importance in
stellar, galactic and cosmological studies.

An accurate census of the small and faint end of the galaxy population is
necessary for the determination of the true galaxy luminosity function,
and is a difficult thing to do \citep{M96,IB97}. 
The Schechter form of the luminosity function with exponent $\alpha$
in the range 1.2-1.8, a popular region, formally diverges.  If there is
not to be an infinite number of faint galaxies, the function must truncate;
this point has not yet been found, and faint galaxies dominate by number
in volume-limited samples \citep{MD97}.  
The luminosity function itself provides strong constraints on theories of
galaxy formation \citep{BST88}.  In particular, the number of dwarfs in
the Local Group appears to be underpredicted
by an order of magnitude in hierarchical
clustering theories \citep{Klyp99}.   Although refinements in theory narrow
the discrepancy \citep{BFL01}, the problem remains (see our discussion
below). 

Apart from numbers, the clustering
properties of dwarfs compared with larger galaxies contains important
information on such things as the bias parameter and the spectrum
of initial density fluctuations from which galaxies grew \citep{BST88}.

Dark matter dominates the internal dynamics of dwarfs 
to a much greater degree than larger objects \citep{M94, H96}, and thus
they provide an especially good place to investigate the nature of this
mysterious stuff.  The external dynamics has been used to derive an
age of the universe independent of the global Hubble constant \citep{L81,
W99} as well as the mass of the Local Group.

The star-formation histories of dwarf galaxies appear to be qualitatively
different from those of larger objects; indeed, within the Local Group
each dwarf differs from each other \citep{H94, G97, M98}.  They thus provide
valuable laboratories to study the general phenomenon of star formation
in a variety of environments.

\section{Dwarf Hunting}

One expects LSB dwarf galaxies to be hard to find.  It requires a great deal
of care to detect an object which is actually fainter than the night sky
(and our targets are typically only a few percent of dark sky brightness).
This means that in any survey they will be seriously underrepresented.
Accordingly, one must look at a lot of sky to find a few (or concentrate
effort where there is likely to be a relative abundance of them). 
For this reason
photographic plates, which provide large area coverage, have been the
popular basis of searches.  However, CCDs provide greater sensitivity and
more controlled enhancement of small signals in large noise, so when
possible they have been used.  (For a discussion of
selection effects affecting LSB objects, see \citet{DI99}, especially the
contributions of Freeman, of McGaugh and of Huchra; also useful is
the survey of surveys by Delcanton in the same volume.)

To correct for the unavoidable incompleteness and bias of a survey it
is useful to have strict criteria, and if possible to perform the survey
automatically.  \citet{A98} and \citet{A99} have used an enhancement
technique on digitized POSSII plates to discover new satellite galaxies
of M31.
However, any direct detection algorithm for low-signal objects tends to
produce a very large number of false detections.  For this reason, and
to cover larger areas in the sky, visual examination of photographic
sky surveys is a viable technique.  \citet{KK98}, \citet{KKS00} and
\citet{KK00} examined film copies of
the POSSII plates (with some success, \citet{KK99}), as did \citet{C97}.

As opposed to limited-area searches (for example, 
around nearby galaxy groups where the probablility of
encountering a galaxy of any type
is assumed to be higher),  objects in the Local Group
might be anywhere in the sky, requiring a full 4$\pi$ steradian
search. In this case 
the requirement for coverage, together with the small physical size of
CCDs, have made the use of photographic methods mandatory for our current
efforts\footnote{
Great efforts are currently being made to expand the area coverage of CCD 
surveys.  The Sloan Digital Sky Survey, to name one, covers a 
significant fraction of the sky.  However, data reduction techniques used
in (at least) the first release of Sloan data
will not allow the detection of large, low surface brightness
objects.  All-sky searches for this type of object are several years away.
We will watch with interest to see whether there is a large
population of extremely low surface brightness objects visible only to
deep CCD surveys.}.
Avoiding also the construction of an elaborate automatic algorithm,
 we decided to perform a visual search of survey plates
starting with the southern sky.

To this end a visual examination of all 894 fields covered by the
ESO-SRC and SERC Equatorial surveys of the southern sky $(\delta < 3\arcdeg)$
was performed.  It was discovered early on that the glass copies of the
surveys were significantly better than film copies for picking up very
faint objects, so glass was used exclusively.
Objects resembling the Andromeda dwarf spheroidals and the
Tucana dwarf,
that is of very low surface brightness, diffuse and large (one to a
few minutes of arc), were noted.  (Very nearby objects, of large angular
size and completely resolved, would tend to get lost among the
star foreground, so our search is not sensitive to close Milky Way satellites.
We had no chance, for instance, of finding the Sagittarius dwarf
visually.  Paradoxically, more distant objects are easier to see.) 
The ``bright'' limit of accepted objects is difficult to quantify
and probably varied significantly from plate to plate, though in all
cases was well above the brightness of the known dwarfs.  The faint
limit, within the variations of the plate material itself, was
probably more stable.  (For an {\em a posteriori} examination of
the limits and reliability of the survey, see section 5.1.)
The size was determined with a ruler and the image
on the plate, which was not always the same size as that found during
followup observations and is in any case ill-defined for most of these
objects.  No candidate measuring one minute or over in all directions was
rejected as being too small; a few cases which appeared to be on the
borderline and are in fact smaller were included.

Even though LSB galaxies cannot be
detected near the galactic plane, {\em all} plates were examined. 
At these signal levels the distribution of Galactic nebulosity dense enough
to interfere with the search is very irregular and difficult to
predict, so we avoided the problem of prediction
altogether.  As a byproduct, we discovered 
several planetary nebulae (which are interesting in their own right).

Plate defects were screened out by
requiring an object's presence on both blue and red plates.  The same
procedure eliminated most reflection nebulae due to their blue colors.
Dwarflike objects noted by the UK Schmidt Telescope Unit during the survey
\citep{T74}, but not previously followed up, were also added to the list.
Catalogs of known objects were consulted (chiefly the NASA Extragalactic
Database, NED) and galaxies of high radial velocity as well as 
planetary nebulae were excluded.  Finally, when time at the telescope
for followup observations permitted, we examined objects which were not
necessarily LSB dwarfs but had been identified by others as possible
Local Group members.

\section{Followup Observations}
 
The majority of the objects from our list (both those derived from survey
plates and those from other sources) were imaged using the 1.5m
telescope at Cerro Tololo Interamerican Observatory (CTIO) with a thinned Tek
$2048 \times 2048$ CCD.  Some of the more northerly ones
were followed up using the central CCD of the Wide Field Camera (WFC) on the
2.5m Isaac Newton Telescope (INT) at the Observatorio del Roque de los
Muchachos in the Canary Islands.  At the f/13.5 Cassegrain focus of the 1.5m
CTIO telescope the scale is 0.24 arcsec/pixel and the field
coverage is just over $8 \times 8$ arcmin.  The central thinned 4k$\times$2k
EEV CCD on the INT WFC, mounted at prime focus, gives a spatial coverage of
$22 \times 11$ arcmin with a scale of 0.33 arcsec/pixel.
 
Candidates were initially examined by taking 15--20 minute exposures
in the R band.  With seeing between 0.9 - 1.4 arcsec this
enables stellar objects to $R \approx 23$ to be detected.  At this depth
objects close to or within the Local Group should begin to resolve into
stars, with the tip of the giant branch becoming readily visible.
If a candidate appeared to resolve into stellar components, further
broadband observations in V and I together with narrowband
${\rm H} \alpha$ were obtained.  The raw CCD frames were processed in
the standard way (bias-subtracted, trimmed and flat-fielded using
twilight flats taken during the observing run) in almost real time at the
telescope to aid in visual inspection of candidates.  A series of
standard star fields taken from \citet{L92} were observed at intervals
throughout each night.

\section{Data on the Dwarf Candidates}

Positions and characteristics of the dwarf galaxy candidates we examined are 
summarized in table \ref{data1}.  We have listed first those objects 
derived from our examination of survey plates, then those taken from other
sources in the literature.  Where the object already had a designation,
we have generally refrained from giving it another; where none existed, we have
constructed one based on IAU recommendations 
(see \url{http://cdsweb.u-strasbg.fr/iau-spec.html}).  KK98 denotes objects
from \citet{KK98}, KKS00 from \citet{KKS00}, and KK00 from \citet{KK00}.
Where more than one designation exists for an object we have tried to list
all of them (in the first column), 
since our experience is that different workers prefer different
catalogs.
The positions are good to within 0.1 minute of arc\footnote{Too much should
not be made of this accuracy.  For
many of these objects, minutes of arc in size and very diffuse, one could
reasonably choose a position for the ``center'' which is a large fraction
of a minute away from that given here.}. 
The size is that of the visible
image at the followup telescope, not of any strictly determined isophote
(and would of course be different between the CTIO 1.5m and the INT 2.5m).
The nature is determined from the R-band morphology, except in the cases
where time allowed an H$\alpha$ image to confirm (or not) emission
nebulae.  Additional comments on each object are presented in the next
section.
Images of all objects are presented
in the figures.  The scale varies somewhat between objects as indicated
in the captions; in all, north is at
the top, east to the left.  In the cases of certain objects (notably
the Antlia and
Cetus dwarfs) CTIO images are presented even though better data are 
available, in the interest of providing a common
comparison with the other objects.

\subsection{Comments on Individual Objects}

{\em ESO 410G005} \\
Although this object is clearly resolving, it is slightly too far away to
be a Local Group galaxy.  \citet{KSG00} place it in the Sculptor Group on the
basis of HST CMD data.

{\em Cetus} \\
The Cetus dwarf spheroidal is larger and much more diffuse than most of
the objects in our list of candidates.  For more detailed information,
see \citet{WHI99}.


{\em WHI B0113-62} \\
Nebulosity connected with the galaxy cluster Abell S0143.


{\em WHI B0200-03} \\
This is clearly resolving even on the survey plates.  The followup images
show it to be a cluster of blue stars (with a distant galaxy cluster in the
background).  

{\em ESO298G033} \\
A galaxy with a starlike nucleus and very faint disk.

{\em PGC 009140} \\
This is a dwarf spheroidal galaxy, beginning to resolve at the faintest
levels (thus outside the Local Group, but not far).

{\em WHI B0240-07} \\
Possibly a LSB barred spiral.  No detectable resolution.

{\em WHI B0441+02} \\
Irregular in shape, possibly spiral.  There is some
granulation which could come from HII regions.

{\em ESO 085G088} \\
This object is either in or behind the Large Magellanic Cloud.  From its
symmetrical, diffuse morphology we are inclined to call it a galaxy, rather than
an LMC nebula.


{\em WHI B0619-07} \\
This appears to be a galaxy showing traces of resolution, though a
superimposed foreground star makes study difficult.

{\em WHI B0652+00} \\
A faint reflection nebula.

{\em PGC 020125} \\
Irregular in morphology,
resolving into stars clearly.  However, the CMD shows visible stars to
be supergiants, placing it at a distance of 2-4 Mpc.  Informally
designated the ``Argo dwarf'' (it lies in Carina, but there is already
a Carina dwarf, so the disused constellation of Argo was resurrected
for naming purposes).

{\em WHI B0713-44} \\
A round, brighter section of a more general Galactic nebulosity.

{\em WHI B0717-07} \\
Extremely faint round object.  An image in H$\alpha$ shows it to be a
smooth planetary.

{\em PGC 020635} \\
Irregular in form, with foreground stars and background galaxies making
detailed study difficult.

{\em ESO368G004} \\
Foreground stars give an illusion of resolution at first glance.

{\em PGC 021406} \\
A faint, diffuse galaxy.

{\em WHI B0740-02} \\
The low-surface brightness object is the extended halo of a cD galaxy
at the center of a cluster.

{\em WHI B0744-05} \\
Possibly spiral, though a foreground star makes it difficult to be sure.

{\em WHI B0750-55} \\
A round, diffuse galaxy.

{\em ESO 165G006} \\
Although cataloged as a galaxy no less than three times, this object is
in fact a rather pretty planetary nebula, as shown by the hydrogen-alpha 
image.

{\em WHI B0921-36} \\
A very faint, large area of Galactic nebulosity.

{\em ESO 126G019} \\
This object has an overall diffuse morphology.  However, at the very lowest
signal level there is structure which may be the beginning of resolution.

{\em KDG 058} \\
Very faint, diffuse galaxy.

{\em WHI B0959-61} \\
A large reflection nebula.  No detectable H$\alpha$.

{\em Antlia} \\
The clear resolution of Antlia into stars provides a contrast with
most other objects in this catalog.  Further details are provided 
in \citet{W97}.


{\em PGC 030367} \\
Extremely large and diffuse.  No detectable H$\alpha$.

{\em WHI B1030-62} \\
A relatively large emission nebula, probably a planetary, as shown by
our H$\alpha$ image.

{\em ESO 215G009} \\
Diffuse galaxy.

{\em WHI B1103-14} \\
Across the north end of this faint, diffuse galaxy lies a chain of HII regions
(confirmed by H$\alpha$ imaging).

{\em WHI B1117-68} \\
A large, dim reflection nebula.

{\em PGC 035171} \\
Faint, diffuse galaxy.


{\em PGC 036594} \\
Very faint galaxy.

{\em WHI B1241-54} \\
Faint reflection nebula.

{\em WHI B1243-20} \\
A large, faint, diffuse bit of Galactic nebulosity.

{\em WHI B1249-33} \\
Very large and diffuse.  Noted by the UKSTU.

{\em ESO 269G066} \\
Diffuse galaxy.

{\em PGC 046680} \\
Diffuse galaxy.

{\em PGC 048001} \\
Galaxy behind a thick layer of foreground stars.

{\em PGC 048178} \\
This galaxy shows traces of a bar and ring structure.
 
{\em ESO174G001} \\
This galaxy is showing some resolution into stars at the limit of
detection, but only after stacking several 
exposures.  It is therefore too distant to be in the Local Group,
but is in the Local Volume.

{\em WHI B1414-52} \\
A planetary nebula, well shown in our H$\alpha$ image.

{\em WHI B1425-47} \\
Faint reflection nebula.
	
{\em WHI B1432-47} \\
The object derived from the survey plate is the ill-defined central
brighter region of a general nebulosity that fills the CTIO 1.5m field.
There is no detectable H$\alpha$ emission.

{\em WHI B1432-16} \\
Several of the knots on the south side of this galaxy are HII regions,
confirmed by an H$\alpha$ image.

{\em WHI B1444-16} \\
A bright, round object, probably a ring galaxy.

{\em WHI B1505-67} \\
A faint wisp, which is probably part of an even fainter complex of
Galactic nebulosity.

{\em WHI B1517-41} \\
A very faint and ill-defined bit of Galactic nebulosity.

{\em WHI B1603-04}\\
Diffuse galaxy.

{\em PGC 057387} \\
Although catalogued as a galaxy, this is clearly a Galactic reflection
nebula.

{\em WHI B1619-67}\\
Faint reflection nebula.

{\em PGC 058179} \\
This is probably an obscured galaxy, though it could possibly be a
Galactic reflection nebula.

{\em WHI B1728-08} \\
A faint ring, shown to be a planetary by the H$\alpha$ image.

{\em WHI B1751-07} \\
Although this looks like a spiral galaxy, there is detectable H$\alpha$
in the same shape, so we identify it as a planetary nebula.


{\em PGC 062147} \\
There are possibly traces of spiral arms in this low surface brightness
galaxy.

{\em ESO 458G011} \\
A small, heavily obscured disk galaxy.

{\em WHI B1919-04} \\
A heavily obscured galaxy.

{\em WHI B1952-04} \\
An obscured galaxy with a prominent bar and ring.

{\em ESO 027G002} \\
Diffuse, slightly elongated.

{\em WHI B2212-10} \\
A face-on disk galaxy, with very faint spiral arms.

{\em ESO 468G020} \\
A dwarf spheroidal galaxy, too distant to be in the Local Group but not
far away.  Clearly resolving.  Although it lies in the general direction
of the Sculptor Group, it is probably too far away to be a dynamical
member.

{\em WHI B2317-32} \\
This is a puzzling object.  On the survey plate it appears to be resolving
into a handful of stars, but our image shows these to be lumps in a more
general nebulosity which is of vaguely spiral form.  An image in H$\alpha$ 
shows that none of these is an HII region, so it is neither a distant
star-forming spiral nor a young Galactic star cluster.  To add to the
mystery, \citet{C97} detected it in HI (their object SC 2) with a 
low heliocentric radial velocity of 68 km s$^{-1}$.  
We tentatively conclude that
the identification of the radial velocity with the galaxy is in error, and
that this is actually a distant galaxy.

{\bf Objects from Other Sources}

{\em SC 18} \\
Included in the Sculptor Group survey of \citet{C97}, though on the basis
of their available data possibly nearer.  Our image, although lumpy, is not
resolving, so it is probably in the Sculptor Group (though not on the near
side).

{\em ESO 293G035} \\
Identified as a dwarf galaxy in the nearby Sculptor Group by \citet{C97}.
Our image shows
a more distant object with a large, low surface brightness plume.  

{\em LGS 2} \\
Identified by \citet{KLS78} as a possible Local Group galaxy (along with LGS 3,
which is such an object).  Our image shows it to be a brighter region
in a large diffuse nebula, so we may answer \citet{VR81} with a confident
``no.'' 

{\em SC 24} \\
Also in the Sculptor Group survey of \citet{C97} and thought to be 
possibly nearer.  Our image is again lumpy, though not resolved, so it
is probably also in the Sculptor Group.

{\em ESO 352G002} \\
Identified as  Local Group galaxy by \citet{SB92} on the basis of radial
velocity.  We find a distant spiral galaxy
with a superposed star (which probably provided the low radial velocity).

{\em A0107+01} \\
Identified as  Local Group galaxy by \citet{SB92} on the basis of radial
velocity.  We find no identifiable galaxy at this position, just a star
(which could have provided a low radial velocity).

{\em ESO 416G012} \\
Identified as  Local Group galaxy by \citet{SB92} on the basis of radial
velocity.  Our image is clearly of a background spiral, with a star 
superposed which probably provided the low radial velocity.  A higher
radial velocity has since been determined.

{\em IC 1947} \\
Identified as  Local Group galaxy by \citet{SB92} on the basis of radial
velocity.  Our image shows a background spiral with several 
neighbors.  The starlike nucleus may in fact be a foreground star; this
would explain the two low radial velocities in the literature.  A
radial velocity consistent with a much larger distance has since been
determined.
   

{\em ESO 056G019} \\
Identified by \citet{SB92} as a Local Group galaxy on the basis of radial
velocity, and by NED as a part of the Large Magellanic Cloud on the same
basis.  Our H$\alpha$ image shows a very pretty emission nebula, with
a bright central part about three minutes across surrounded by a faint
shell about eight minutes in diameter.

{\em ESO 318G013} \\
Identified by \citet{SB92} as a Local Group galaxy on the basis of radial
velocity.  Our image (R, V and I combined) shows a lumpy spindle, possibly
beginning to resolve; however, a H$\alpha$ image shows that the brightest
of the knots are HII regions.  It is therefore well outside the Local
Group.  A superimposed star is probably responsible for a misleading
radial velocity.  (The bright H$\alpha$ regions should provide a good
optical radial velocity for any observers who care to take up the challenge.)

{\em ISI 1106+0257} \\
This galaxy was chosen from \citet{ISI96} 
for its low radial velocity (-69 km/s).  (Note that those authors make
no claim as to Local Group membership.)  Our picture shows a faint barred
spiral with a superimposed star; the latter is probably the source of
the radial velocity.

{\em ESO 269G070} \\
Identified by \citet{SB92} as a Local Group galaxy based on radial velocity.
Our image shows a smooth disk galaxy with a superimposed star, which is
probably the source of a misleading radial velocity (a more believable
velocity has since been determined).  There is a faint
companion galaxy almost in contact to the west.

{\em IC 4247} \\
Identified by \citet{SB92} as a Local Group galaxy based on radial velocity.
Our combination of three images (total of one hour exposure) shows just the
beginning of resolution.  It is too far away to be in the Local Group, but
is in the Local Volume.
 
{\em NGC 5237} \\
Identified by \citet{Fa81} as a member of the Local Group based on radial
velocity.  Interestingly, identified as a member of the  NGC 5128
(Centaurus) Group by \citet{HH84} on the same basis.  Our image shows
a smooth distribution of light with no trace of resolution, which places
it well beyond the Local Group.

{\em IC 4739} \\
Identified as a Local Group galaxy by \citet{SB92} on the basis of radial
velocity.  Our image clearly shows a distant spiral.  A superimposed star
is probably the source of an erroneous radial velocity, since corrected
in the literature.

{\em IC 4789} \\
Identified as a Local Group galaxy by \citet{SB92} on the basis of radial
velocity.  Our image clearly shows a distant spiral with a superposed
star, the latter probably providing a misleading radial velocity (since 
corrected in the literature).

{\em IC 4937} \\
Identified as  Local Group galaxy by \citet{SB92} on the basis of radial
velocity.  Our image clearly shows a distant edge-on spiral.  A superposed
star is probably responsible for a misleading radial velocity.

{\em IC 5026} \\
Identified as  Local Group galaxy by \citet{SB92} on the basis of radial
velocity.  Our image clearly shows a distant edge-on spiral.  A superposed
star is probably responsible for an anomalously low radial velocity.

{\em A2259+12} \\
Identified as  Local Group galaxy by \citet{SB92} on the basis of radial
velocity.  We find a small, bright galaxy with
no trace of resolution.

{\em ESO 347G008} \\
Identified as  Local Group galaxy by \citet{SB92} on the basis of radial
velocity.  Our image shows a very diffuse object, not resolving into stars.
It is possible that very faint outlying lumps actually trace spiral
arms; these (as well as two brighter objects near the center) show up
as well in the H$\alpha$ image.  If this galaxy is indeed a very low
surface brightness face-on spiral, the size is
actually something like 2.6 arc minutes.  A higher radial velocity has
since been determined.

\section{Summary and Discussion}

\subsection{Completeness of the Local Group Census}
This portion of the Dwarf Hunt was successful in its primary object,
to find new Local Group dwarfs\footnote{We count Antlia as a Local Group
object for convenience, even though it might not be gravitationally
bound to the Group.  Observationally it shares the characteristics of undisputed
members, such as resolution into Giant Branch stars.}, as well as its
secondary object, to get some idea of how many Local Group galaxies
might remain to be found.  Unfortunately, the completeness of the survey
is difficult to quantify, for three reasons: the sensitivity of the
survey plates varies with weather and the condition of the photographic
emulsion; galactic extinction varies in a very complicated way with
position at low signal levels; and the response of the eye to faint
signals is complex and depends on many things.  We consider each
of these effects in turn below.

\citet{MES90} studied the variation in sensitivity among the 185 survey
plates used for the APM Galaxy Survey.   They concluded that the rms
plate-to-plate variation amounted to 0.178 magnitude, and that 
vignetting amounted to 20\% loss of light as far as 2.5$^{\rm o}$
from the center of the field.  Combining these we may estimate
0.3 magnitudes of variation due to plate material.  Since the central
(brightest) region of Cetus is 25.05 magnitudes per square arc
second (and it is not the faintest object in our list) we very
conservatively conclude
that our survey was complete to at least 24.7 mag arcsec$^{-2}$ in regions
clear of Milky Way interference. 
In particular, we went deep enough to see the majority of very faint
objects in the simulations of \citet{BFL01}. 

An estimate of the fraction of dwarf galaxies hidden behind the
Milky Way may be made using a grading system instituted while searching the
plates.  Each plate was noted as ``good'' (424 plates total),
``troublesome'' (315 plates)
or ``difficult'' (158) based on the amount of background nebulosity
interfering with detection of faint galaxies.  Taking as a rough measure
that no dwarfs will be seen on ``difficult'' plates, none over about
half the area of ``troublesome'' plates, and that any dwarf would be seen on
a ``good'' plate, about 35\% of the sky could contain an unseen faint
dwarf galaxy.  With eleven of this type of galaxy known in the southern
sky (Stextans A, WLM, Fornax, SagDIG, Sculptor, Sextans, Phoenix, Tucana,
Carina, Antlia and Cetus), we would expect something like four more to
be hiding behind the Milky Way.  This estimate may be somewhat low, since
nebulosity which obscures but is not luminous would not have affected
the plate grading; but this is probably not a big effect.
(The interesting,
though not Local Group, galaxy Argo was found on a ``troublesome'' plate).

\subsubsection{Reliability of Visual Surveys}

There remains the hardest quantity to estimate: the reliability of
visual processing of the images.  The complicated nature of visual
response makes a calculation from theory impractical; the only
useful estimate must come from a comparison of results of independent
examinations of the same data set.

The most comparable survey to ours is that reported in \citet{KK98}, \citet{KKS00}
and \citet{KK00}.  Both surveys covered the entire southern sky by
visual examination of survey photographs.  The object of the other
group was different, however: to find dwarf galaxy candidates in the
entire Local Volume, well beyond the Local Group.  They thus included
objects smaller than our limit and applied slightly different standards
of morphology.  To a first approximation, however, any object meeting
our standards also met theirs.  The
two surveys are independent, since neither used the results of the other
in compiling their lists.

Our first task is to estimate $N$, the total number of objects 
visible on the southern survey plates meeting criteria common to both
searches: of very low surface brightness, one to a few minutes
of arc in size, not obviously ruled out by morphology 
nor by high radial velocity.
Of these, our survey found $n_1$ and the Karachentsev group $n_2$.
The probabilies of a given object being found by each group are
then $n_1/N$ and $n_2/N$, respectively.
The number missed by us but found by the other group, $m_1$, is the
probability of our missing an object, times their probability of 
finding it, times the number of objects:
\begin{equation}
m_1 = \left( 1-\frac{n_1}{N} \right) \frac{n_2}{N} N
\end{equation}
So our estimate of the total number of objects is
\begin{equation}
N = \frac{n_1 n_2}{n_2 - m_1}
\label{eq:N}
\end{equation}
The next step is to determine these numbers.

Twenty-two objects were located by both groups.  There are 29
objects found by the other group which they report as
large enough (over a minute
of arc in both dimensions) to fit our criteria and yet do not appear
in our list.  Of these, a review of Digital Sky Survey images and film copies
of the Survey shows 12
too bright for us to record\footnote{We think it reasonable to assume
that a galaxy with relatively high surface brightness is a big galaxy
far away, based on the morphology of known members of the Local Group.
``Far away'' in this context may mean within the Local Volume, of course,
so the Karachentsev \& Karachentseva group did not apply the same
restriction.}.  A further five were rejected by us because their
known radial velocities were too high for membership in the Local Group.  
Six are morphologically arguable, in that they didn't look like dwarf
galaxies to us but did to the Karachentsev group.
This leaves six objects meeting
our selection criteria which we missed; add to these, say, half of the
morphologically arguable objects, for a total of nine.

On the other hand, there are 37 objects in our survey not noted by the other
group: 21 galaxies, ten reflection nebulae and six planetaries (leaving
out the star cluster and the galaxy clusters).  Pending a close comparison
of search techniques and criteria no clear conclusions can be drawn about
why these were missed.  On the basis of our experience with survey films
as opposed to plates, we suggest that some of the explanation
might be due to the lower noise
levels of the latter, allowing us to detect fainter objects.  It is also
probable that many, or all, of the 16 galactic objects were rejected on
morphological grounds.

We use equation~\ref{eq:N} to calculate two cases: in the first, 
the non-galaxy objects missing
from the Karachentsev group's listing but found by us are excluded (which
would be the case, for example, if their morphological selection were
perfect for these objects); in
the second they are included.  Thus $n_1$ is 43 or 59, $n_2$ is 31, and
$m_1$ is 9.  These result in a total number of
objects of 61 and 83, respectively, of which 43 and 59 have been 
found\footnote{We exclude the nine objects missing from our list because
(as far as we know) they havn't yet been shown {\em not} to be Local Group
dwarfs.}, leaving 18 or 24 objects still lurking on the plates.

Consider now that of the 43 or 59 objects examined, two have proven to be
Local Group dwarfs, giving a conversion rate of .047 and .034.  In each
case we thus expect eight-tenths of a Local Group dwarf to remain among
the objects on the plates.  

Our overall estimate of Local Group dwarf galaxies remaining undiscovered
in the southern sky, either hidden by the Milky Way or missed by two
visual surveys, is thus about five.  

Although some of the assumptions and numbers going into our numerical
estimate are uncertain or arguable, the overall qualitative conclusion
is very robust.  Let us assume, for example, a very high conversion rate
of faint objects to Local Group galaxies by including all eleven known
in the south and dividing by the low estimate of 43 objects plus
the nine known previously (which were not in our list).  With this
conversion rate of about one in five, the number of additional
objects required to
double the Local Group's census, say 36, comes out to 180.  For this number
of objects to exist on the plates and still remain unseen our survey
would have had to miss almost three times as many objects as it found,
of exactly the type we were looking for.  The overlap with the other
group's list would be expected to be
about seven or eight, one-third of the actual
overlap. And this is still not enough of an expansion of Local
Group membership to agree with cosmological simulations.

Our estimate of completeness of course only takes into account those 
objects which fall within our search criteria.  We think it quite
reasonable to assume 
that all large, bright galaxies in the Local Group (outside
the region hidden behind the Milky Way) have already been found.
Our search was sensitive to dwarf galaxies as faint as any known.
Nearby dwarfs could have remained hidden, if they were close and
diffuse enough for their stars to be lost among the Milky Way foreground;
however, the volume of space available for this is not big enough
to hide many of them.  Galaxies that are much more concentrated than
any known, so that they fell below our size criterion; or conversely
much less concentrated, so that their surface brightness fell below
the plate limits, would be missed.  Objects that did not have Giant
Branch stars might not be seen, but that requires no star formation
at all, or a strange sort which is not elsewhere seen.  

\subsection{Other Objects in the Catalog}

There are a large number of Galactic reflection
nebulae in our list.  We attempted
to avoid them by requiring any object to appear on both colors of 
survey plate, assuming that (blue) nebulae which are very faint
on the blue plates would not appear at all on red plates.
This is a reasonable assumption \citep{WS86} and surely many
were removed this way.  However, apparently there are a range of
colors among this kind of object and several were bright enough in
the red to make it through.

Somewhat surprising are the galaxy clusters we found.  These
were not detected on the plate due to the concentration of galaxies, but
rather due to a faint glow.   Although this is generally centered on a
dominant galaxy, it is very spread out and probably consists of 
intracluster stars not bound to any one galaxy.

As expected, several planetary nebulae were found near the Galactic plane.
From their large apparent size and very low surface brightness, these must
be either very old or heavily obscured (and in either case, rather close).
Old planetaries are of particular interest due to their interaction
with the interstellar medium  \citep{BSS90}, though
in these cases the very low surface brightness will make further study
by spectroscopy difficult.

The majority of our objects are galaxies far beyond the Local Group.
These come in a variety of shapes, with diffuse dwarfs predominating but
face-on disk galaxies well-represented.  In many cases the nuclei of the
latter are actually rather bright, but appeared starlike on the survey
plates and so were not obviously extragalactic.

In addition to the list of objects derived from our own examination of
survey plates, we had the opportunity to investigate several which were
suggested by other workers as possible or probable Local Group galaxies.
None of these turned out to be in the Local Group (though a few are
probably nearby).  The basis for almost all of these suggestions was radial
velocity; either a low value was taken to indicate Local Group membership,
or a more elaborate model (as in \citet{SB92}) was applied. 
Both procedures are invalid, of course, when the velocity data point is
incorrect, as it is for most of these.  

This fact should not be taken
as indicating a high level of error in radial velocity catalogs.  
The few instances highlighted here of superimposed stars giving misleading
velocities are taken from many thousands of measurements, and most of
these have been corrected.  (Superimposed Galactic stars will, of course,
only give erroneously {\em low} radial velocities to distant galaxies.)

We do wish to point out, however, the danger of forming conclusions based
on a single type of data.  Errors will always creep in, and if 
there is no independent method of detecting them they will remain.  
Astronomy (especially cosmology) is necessarily dependent on datasets
containing sparse data on many objects and so is particular subject
to this problem.  We urge, whenever possible, the combination of different
types of data.  In what is no doubt an extreme example, by
simple imaging we have reduced the population assigned by \citet{SB92}
to the Local Group by roughly 20\%.

A similar problem shows up in the cases of NGC 5237 and ESO 056G019.  In  
each, the same radial velocity was taken to indicate quite different
locations by different authors.  Again, imaging was able to resolve the
disagreement. 

\subsection{Further Investigations}

The obvious next step in the Dwarf Hunt is to extend the search to the
other half of the sky.
A similar visual survey of the northern sky is now being performed using
POSSII plates, and we expect to report the results from it in due course.
In addition, the objects from the Karachentsev \& Karachentseva group
which conform to our criteria but did not appear in our list should be
followed up.

Meanwhile, the objects here presented offer much scope for further study.
Lines of inquiry from the Galactic (interaction of old planetary nebulae
with the interstellar medium) to the cosmological (formation and evolution
of large, distant low surface brightness galaxies) are represented.  In
particular there is useful work to be done on the resolved dwarf
galaxies in the Local Volume. 

\acknowledgements
The authors wish to thank John Pilkington of the now defunct
Royal Greenwich Observatory
for digitally scanning many survey plates and Sue Tritton of the Royal
Observatory Edinburgh for her patient assistance in distinguishing faint
astronomical objects from plate defects.  
The authors are grateful for partial financial support from the Institute
of Astronomy of the University of Cambridge and the Physics Department
of the U. S. Naval Academy.  GKTH acknowledges financial support from
Chilean FONDECYT grant 1990442.  In addition, we would like to express our
appreciation for the dedicated help of the night assistants at Cerro Tololo.
This research has made use of the NASA/IPAC Extragalactic Database
(NED) which is operated by the Jet Propulsion Laboratory, California
Institute of Technology, under contract with the National Aeronautics
and Space Administration.

\clearpage

\begin{deluxetable}{lcccc}
\rotate
\tablewidth{0pc}
\tablecaption{Dwarf Galaxy Candidates}
\tablehead{
\colhead{Designation(s)} & \colhead{RA (HH-MM-SS)} & \colhead{Dec (DD-MM-SS)}
& \colhead{Size (arc min)} & \colhead{Type} }
\startdata
\sidehead{Plate Survey Dwarf Candidates (Positions for 1950.0):}
 ESO 410G005 & 00 13 00.3 & -32 27 28 & 1.7 $\times$ 1.1  & nearby galaxy \\*
ESO 001300-3227.6 & & & & \\*
UKS 0013-324 & & & & \\*
AM 0013-322 & & & & \\*
PGC 001038 & & & & \\ *
KK98-3 & & & & \\
Cetus dwarf spheroidal & 00 23 39 & -11 19 16 & 5 & Local Group galaxy\\*
WHI B0023-11 & & & & \\ *
KKS00-1 & & & & \\
WHI B0113-62 & 01 13 41.1 & -62 31 54 & 1.4 $\times$ 0.7 & galaxy cluster \\*
Abell S0143 & & & & \\
WHI B0200-03 & 02 00 25.1 & -03 29 34 & 1.2 $\times$ 1.0 & star cluster \\
ESO 298G33 & 02 19 26.0 & -39 01 54 & 1.3 $\times$ 0.6 & galaxy \\*
ESO 021926-3901.9 & & & & \\*
ESO-LV 298033 & & & & \\*
AM 0219-390 & & & & \\
PGC 009140 & 02 24 25.0 & -73 44 19 & 2.7 $\times$ 1.4 & nearby galaxy \\*
SGC 0224.3-7345 &  &  & &  \\* 
KK00-3 & & & & \\
WHI B0240-07 & 02 40 09.9 & -07 32 59 &  0.7 $\times$ 0.9 & galaxy \\
WHI B0441+02 & 04 41 07.6 & +02 54 00 & 0.9 & galaxy \\
ESO 085G88 & 05 26 47.0 & -63 16 54 & 1.7 $\times$ 0.6 & galaxy? \\*
ESO 052647-6316.9 & & & & \\*
ESO-LV 0850880 & & & & \\*
SGC 052647-6316.9 & & & & \\*
KK00-15 & & & & \\
WHI B0619-07 & 06 19 49.2 & -07 48 51 & 1.5 $\times$ 0.8 & nearby galaxy? \\
WHI B0652+00 & 06 52 02.4 & +00 18 49 & 1.5 & reflection nebula \\
PGC 020125 & 07 04 29.7 & -58 26 33 & 1.4 $\times$ 1.9 & nearby galaxy \\*
SGC 0704.7-5826 & & & & \\*
AM 0704-582 & & & & \\*
Argo dwarf irregular & & & & \\
WHI B0713-44 & 07 13 28.6 & -44 18 41 & 1.0 & reflection nebula \\
WHI B0717-07 & 07 17 14.4 & -07 07 35 & 1.0 & planetary nebula \\
PGC 020635 & 07 17 41.1 & -57 19 07 & 1.6 $\times$ 1.1 & galaxy \\*
SGC 0717.6-5718 & & & & \\*
AM 0717-571 & & & & \\*
KK98-59 & & & & \\
ESO 368G004 & 07 31 05.0 & -35 22 48 & 0.5 $\times$ 0.7 & galaxy \\*
ESO 073105-3522.8 & & & & \\*
KK00-22 & & & & \\
PGC 021406 & 07 37 20.0 & -69 13 38 & 0.9 $\times$ 1.1 & galaxy \\*
AM 0737-691 & & & & \\*
SGC 0737.7-6912 & & & & \\*
KK98-63 & & & & \\
WHI B0740-02 & 07 40 49.9 & -02 25 03 & 0.6 & galaxy cluster \\
WHI B0744-05 & 07 44 16.1 & -05 39 51 & 1.1 $\times$ 0.5 & galaxy \\
WHI B0750-55 & 07 50 15.3 & -55 19 27 & 0.9 & galaxy \\*
KK00-24 & & & & \\
ESO 165G006 & 08 52 09.6 & -53 53 41 & 2.2 & planetary nebula \\*
ESO 085210-5353.5 & & & & \\*
SGC 085210-5353.5 & & & & \\
WHI B0921-36 & 09 20 58.0 & -36 12 57 & 4.7 $\times$ 2.2 & reflection nebula \\
ESO 126G019 & 09 32 52.2 & -61 03 34 &  2.3 $\times$ 2.0 &  galaxy \\*
ESO 093253-6103.4 & & & & \\*
SGC 093253-6103.4 & & & & \\*
IRAS 093253-6103 & & & & \\
KDG 058 & 09 37 53.1 & +00 16 12 & 0.8 $\times$ 0.6 & galaxy \\
WHI B0959-61 & 09 59 00.1 & -61 54 39 & 4.6 $\times$ 3.5 & reflection nebula \\
Antlia dwarf spheroidal & 10 01 47 & -27 05 15 & 1.8 $\times$ 1.3 & Local Group galaxy \\*
PGC 029194 & & & & \\*
SGC 1001.9-2705 & & & & \\*
AM 1001-270 & & & & \\
PGC 030367 & 10 20 13.3 & -32 52 29 &  5.5 & galaxy \\*
SGC 1020.1-3253 & & & & \\
WHI B1030-62 & 10 30 31.3 & -62 54 52 & 4.3 & planetary nebula\\
ESO 215G009 & 10 55 16.0 & -47 54 36 & 3.8 $\times$ 3.1 & galaxy \\*
ESO 105516-4754.6 & & & & \\*
SGC 105516-4754.6 & & & & \\*
ESO-LV 2150090 & & & & \\*
KK00-40 & & & & \\
WHI B1103-14 & 11 03 41.7 & -14 08 07 & 2.4 $\times$ 1.3 & galaxy \\*
KKS00-23 & & & & \\
WHI B1117-68 & 11 17 39.2 & -68 48 54 & 0.4 $\times$ 0.3 & reflection nebula \\*
KK00-41 & & & & \\
PGC 035171 & 11 24 09.3 & -72 20 17 &  0.6 & galaxy \\*
SGC 1124.8-7221 & & & & \\*
KK00-42 & & & & \\
PGC 036594 & 11 42 20.1 & +02 26 31 & 0.8 $\times$ 0.4 & galaxy \\
WHI B1241-54 & 12 41 43.6 & -54 09 02 & 2.1 $\times$ 1.3 & reflection nebula \\
WHI B1243-20 & 12 43 07.0 & -20 15 12 & 5.3 $\times$ 3.4 & reflection nebula \\
WHI B1249-33 & 12 49 07.2 & -33 15 26 & 3.7 & galaxy \\
ESO 269G066 & 13 10 15.0 & -44 37 30 & 1.3 $\times$ 0.9 & galaxy \\*
ESO 131015-4437.5 & & & & \\*
SGC 131015-4437.5 & & & & \\*
AM 1310-443 & & & & \\*
KK98-190 & & & & \\
PGC 046680 & 13 19 06.8 & -42 16 20 & 2.5 $\times$ 1.9 & galaxy \\*
SGC 1319.1-4216 & & & & \\*
KK98-197 & & & & \\
PGC 048001 & 13 32 58.0 & -56 17 04 & 1.4 $\times$ 1.2 & galaxy \\*
SGC 1332.9-5616 & & & & \\
PGC 048178 & 13 34 55.7 & -56 13 25 & 1.6 $\times$ 1.0 & galaxy \\*
SGC 1334.9-5613 & & & & \\*
IRAS 13349-5613 & & & & \\*
KK00-56 & & & & \\
ESO 174G001 & 13 44 45.0 & -53 05 54 & 2.7 $\times$ 0.7 & nearby galaxy \\*
ESO 134445-5305.9 & & & & \\*
SGC 134445-5305.9 & & & & \\*
KK00-59 & & & & \\
WHI B1414-52 & 14 14 08.1 & -52 12 29 & 1.7 $\times$ 1.5 & planetary nebula \\
WHI B1425-47 & 14 25 03.5 & -47 13 53 & 2.2 $\times$ 2.7 & reflection nebula \\
WHI B1432-47 & 14 32 30.9 & -47 46 04 &  3? &  reflection nebula \\ 
WHI B1432-16 & 14 32 38.8 & -16 56 58 & 1.3 $\times$ 0.5 & galaxy \\*
KKS00-47 & & & & \\
WHI B1444-16 & 14 44 13.2 & -16 44 44 & 0.8 & galaxy\\
WHI B1505-67 & 15 06 00.7 & -67 45 00 & 0.5 $\times$ 0.3 & reflection nebula \\*
KK00-61 & & & & \\
WHI B1517-41 & 15 17 45.4 & -41 38 43 & 5? & reflection nebula \\ 
WHI B1603-04 & 16 03 02.4 & -04 26 16 & 1.0 $\times$ 0.6 &  galaxy \\*
KKS00-48 & & & & \\
PGC 057387 & 16 06 03.3 & -65 37 03 & 1.8 $\times$ 0.9 & reflection nebula \\*
SGC 1605.9-6537 & & & & \\*
KK00-64 & & & & \\
WHI B1619-67 & 16 20 02.4 & -67 00 47 & 2.6 $\times$ 1.7 & reflection nebula \\
PGC 058179 & 16 22 59.4 & -60 20 53 & 1.2 $\times$ 0.8 & galaxy? \\*
SGC 1623.0-6021 & & & & \\*
KK98-241 & & & & \\
WHI B1728-08 & 17 28 45.6 & -08 17 02 & 0.9 & planetary nebula \\
WHI B1751-07 & 17 51 07.4 & -07 02 46 & 1.0 $\times$ 0.5 & planetary nebula \\
PGC 062147 & 18 33 06.3 & -57 28 25 & 1.5 $\times$ 0.7 & galaxy \\*
SGC 183307-5727.4 & & & & \\
ESO 458G011 & 18 56 20.0 & -31 16 54 & 0.5 $\times$ 0.4 & galaxy \\*
ESO 185620-3116.9 & & & & \\
WHI B1919-04 & 19 19 23.3 & -04 17 46 & 0.8 & galaxy \\
WHI B1952-04 & 19 53 01.2 & -04 31 45 & 0.7 & galaxy\\
ESO 027G002 & 21 46 06.0 & -80 48 30 & 0.9 $\times$ 0.4 & galaxy \\*
ESO 214606-8048.5 & & & & \\*
SGC 214606-8048.5 & & & & \\*
ESO-LV 0270020 & & & & \\
WHI B2212-10 & 22 12 47.1 & -10 43 36 & 1.0 & galaxy \\
ESO 468G020 & 22 37 56.3 & -31 03 41 & 1.4 $\times$ 0.8 & nearby galaxy \\*
ESO 223756-3103.6 & & & & \\*
SGC 223756-3103.6 & & & & \\*
AM 2237.310 & & & & \\*
KK98-258 & & & & \\
WHI B2317-32 & 23 17 53.6 & -32 10 58 & 0.5 $\times$ 0.6 & galaxy  \\*
SC2 &&&& \\
\sidehead{Dwarf Candidates From Other Sources (Positions for 2000.0):}
SC18 & 00 00 57.7 & -41 09 17.9 & 0.6 $\times$ 0.5 & galaxy \\
ESO 293G035 & 00 06 51.6 & -41 50 22.8 &  0.9 $\times$ 0.5 & galaxy \\
LGS 2 & 00 29 16.2 & +33 20 46 & 2.9 $\times$ 1.6 & reflection nebula \\
SC24 & 00 36 38.9 & -32 34 29.8 & 0.8 $\times$ 0.4 & nearby galaxy \\
ESO 352G002 & 01 04 30.4 & -33 39 16.1 & 0.7 $\times$ 0.5 & galaxy \\
A0107+01 & 1 09 58.4 & +02 07 57.8 &  & star \\
ESO 416G012 & 02 43 36.3 & -31 56 19.6 & 1.8 $\times$ 0.7 & galaxy \\
IC 1947 & 03 30 32.7 & -50 20 20.2 & 0.9 & galaxy \\
ESO 056G019 & 4 53 18.5 & -70 35 55.2 & 8 & emission nebula \\
ESO 318G013 & 10 47 41.9 & -38 51 15.4 & 1.7 $\times$ 0.4 & nearby galaxy \\
ISI 1106+0257 & 11 09 20.4 & +02 40 56.6 & 0.5 & galaxy \\
ESO 269G070 & 13 13 28.2 & -43 22 59.4 & 1.0 $\times$ 0.3 & galaxy \\
IC 4247 & 13 26 44.4 & -30 21 44.7 & 1.2 $\times$ 0.6 & nearby galaxy \\
NGC 5237 & 13 37 41.3 & -42 49 06.4 & 1.2 $\times$ 1.0 & galaxy \\
IC 4739 & 18 40 51.3 & -61 54 06 & 0.9 $\times$ 0.6 & galaxy \\
IC 4789 & 18 56 21.8 & -68 34 11.0 & 1.7 $\times$ 0.7 & galaxy \\
IC 4937 & 20 05 17.8 & -56 15 27.0 & 2.5 $\times$ 0.5 & galaxy \\
IC 5026 & 20 48 26.9 & -78 03 58.7 & 2.5 $\times$ 0.4 & galaxy \\
A2259+12 & 23 01 25.9 & +12 44 07.4 & 0.5 $\times$ 0.3 & galaxy \\
ESO 347G008 & 23 20 49.1 & -41 43 51 & 1.2 $\times$ 0.6 &  galaxy \\*
\enddata
\label{data1}
\end{deluxetable}

\clearpage

\figcaption[f1.ps]{ESO 410G005, two 1200s exposures combined,
CTIO 1.5m in R.  Approx. 4 arc minutes square. \label{ESO410G005}}

\figcaption[f2.ps]{WHI B0023-11, the Cetus dwarf spheroidal,
two 1200s exposures combined, CTIO 1.5m in R.  Approx. 8 arc
minutes square.  \label{WHIB0023-11}}

\figcaption[f3.ps]{WHI B0113-62, 1200s exposure CTIO 1.5m in R.
Approx. 4 arc minutes square. \label{WHIB0113-62}}

\figcaption[f4.ps]{WHI B0200-03, 1200s exposure CTIO 1.5m in V.
Approx. 4 arc minutes square. \label{WHIB0200-03}}

\figcaption[f5.ps]{ESO298G033, 1200s exposure, CTIO 1.5m in R.
Approx. 4 arc minutes square. \label{ESO298G033}}

\figcaption[f6.ps]{PGC 009140, 2 $\times$ 1200s exposure in V combined
with one 1200s exposure in R, CTIO 1.5m.  Approx. 4 arc minutes
square. \label{PGC009140}}

\figcaption[f7.ps]{WHI B0240-07, 1200s exposure CTIO 1.5m in R.
Approx. 4 arc minutes square. \label{WHIB0240-07}}

\figcaption[f8.ps]{WHI B0441+02, four 1200s exposures combined,
 CTIO 1.5m in R.
Approx. 4 arc minutes square. \label{WHIB0441+02}}

\figcaption[f9.ps]{ESO085G088, 1200s exposure, CTIO 1.5m in R.
Approx 4 arc minutes square. \label{ESO085G088}}


\figcaption[f10.ps]{WHI B0619-07, one 1200s exposure each in R, V
and I, CTIO 1.5m.
Approx 4 arc minutes square. \label{WHIB0619-07}}

\figcaption[f11.ps]{WHI B0652+00, 1200s exposure CTIO 1.5m in R.
Approx. 4 arc minutes square.  \label{WHIB0652+00}}

\figcaption[f12.ps]{PGC 020125, three 1200s exposures (one each in V, 
R and I) combined.  Approx. 4 arc minutes square. \label{PGC020125}}

\figcaption[f13.ps]{WHI B0713-44, 1200s exosure CTIO 1.5m in R.
Approx. 4 arc minutes square. \label{WHIB0713-44}}

\figcaption[f14.ps]{WHI B0717-07, 1800s exposure CTIO 1.5m in H$\alpha$.  
Approx.  4 arc minutes square.  \label{WHIB0717-07}}

\figcaption[f15.ps]{PGC020635, four 1200s exposures combined,
 CTIO 1.5m in R.  Approx.
4 arc minutes square. \label{PGC020635}}

\figcaption[f16.ps]{ESO368G004, one 1200s exposure each in V, R and
I combined, CTIO 1.5m.  Approx.
4 arc minutes square. \label{ESO368G004}}

\figcaption[f17.ps]{PGC021406, 1200s exposure CTIO 1.5m in R.  Approx.
4 arc minutes square.  \label{PGC021406}}

\figcaption[f18.ps]{WHI B0740-02, 1200s exposure CTIO 1.5m in R.
Approx. 2 arc minutes square.  \label{WHIB0740-02}}

\figcaption[f19.ps]{WHI B0744-05, 1200s exposure CTIO 1.5m in R.  Approx.
4 arc minutes square. \label{WHIB0744-05}}

\figcaption[f20.ps]{WHI B0750-55, 1200s exposure CTIO 1.5m in R.  Approx.
4 arc minutes square. \label{WHIB0750-55}}

\figcaption[f21.ps]{ESO 165G006, 1200s exposure CTIO 1.5m in R.  Approx.
4 arc minutes square. \label{165G006R}}

\figcaption[f22.ps]{ESO 165G006, 1800s exposure in H$\alpha$ divided by
1200s exposure in R, CTIO 1.5m.  Approx. 4 arc minutes square. \label{
ESO165G006}}

\figcaption[f23.ps]{WHI B0921-36, 1200s exposure CTIO 1.5m in R.  Approx.
8 arc minutes square. \label{WHIB0921-36}}
 
\figcaption[f24.ps]{ESO 126G019, 1200s exposure CTIO 1.5m in R.  Approx.
4 arc minutes square. \label{ESO163G019}}

\figcaption[f25.ps]{KDG 058, 1200s exposure CTIO 1.5m in R.  Approx.
4 arc minutes square. \label{KDG058}}
 
\figcaption[f26.ps]{WHI B0959-61, 1200s exposure CTIO 1.5m in R.  Approx.
8 arc minutes square. \label{WHIB0959-61}}

\figcaption[f27.ps]{Antlia; two 1200s exposures in V, plus one each in
R and I, combined.  Approx. 4 arc minutes square.  \label{Antlia}}


\figcaption[f28.ps]{PGC 030367, 1200s exposure CTIO 1.5m in R.  Approx.
8 arc minutes square. \label{PGC030367}}

\figcaption[f29.ps]{WHI B1030-62, 1800s exposure CTIO 1.5m in 
H$\alpha$. Approx.  8 arc minutes square. \label{WHIB1030-62Ha}}

\figcaption[f30.ps]{ESO 215G009, 1200s exposure CTIO 1.5m in R.  Approx.
8 arc minutes square. \label{ESO215G009}}

\figcaption[f31.ps]{WHI B1103-14, three 1200s exposures combined, CTIO
1.5m in R.  Approx. 4 arc minutes square. \label{WHIB1103-14}}

\figcaption[f32.ps]{WHI B1117-68, 1200s CTIO 1.5m in R.  Approx. 4 arc
minutes square.  \label{WHIB1117-68}}
 
\figcaption[f33.ps]{PGC 035171, 1200s CTIO 1.5m in R.  Approx. 4 arc
minutes square. \label{PGC035171}}


\figcaption[f34.ps]{PGC 036594, 1200s exposure CTIO 1.5m in R.  Approx.
4 arc minutes square. \label{PGC036594}}

\figcaption[f35.ps]{WHI B1241-54, 1200s CTIO 1.5m in R.  Approx. 8
arc minutes square.  \label{WHIB1241-55}}

\figcaption[f36.ps]{WHI B1243-20, 1200s exposure CTIO 1.5m in R.
Approx. 8 arc minutes square.  The object is just below center, 
extending off the frame to the right.  \label{IRASF12429}}

\figcaption[f37.ps]{WHI B1249-33, 1200s CTIO 1.5m in R.  Approx. 8
arc minutes square. \label{WHIB1249-33}}

\figcaption[f38.ps]{ESO 269G066, 1200s CTIO 1.5m in R.  Approx.
4 arc minutes square. \label{ESO269G066}}

\figcaption[f39.ps]{PGC 046680, 1200s CTIO 1.5m in R. Approx. 
4 arc minutes square. \label{PGC046680}}

\figcaption[f40.ps]{PGC 048001, 1200s exposure, CTIO 1.5m in R.
Approx. 4 arc minutes square.  \label{PGC048001}}

\figcaption[f41.ps]{PGC 048178, 1200s exposure CTIO 1.5m in R.
Approx. 4 arc minutes square. \label{PGC048178}}

\figcaption[f42.ps]{ESO174G001, three 1200s exposures combined,
 CTIO 1.5m in R.  Approx.
4 arc minutes square. \label{ESO174}}

\figcaption[f43.ps]{WHI B1414-52, 1800s exposure CTIO 1.5m in H$\alpha$.
Approx. 4 arc minutes square.  \label{WHIB1414-52}}

\figcaption[f44.ps]{WHI B1425-47, 1200s exposure CTIO 1.5m in R.
Approx. 8 arc minutes square. \label{WHIB1425-47}}

\figcaption[f45.ps]{WHI B1432-47, 1200s exposure CTIO 1.5m in R.
Approx. 8 arc minutes square.  The crescent-shaped object to the
upper left is a reflection within the optical system.  WHI B1432-47
is the ill-defined central brighter part of the general nebulosity.
\label{WHIB1432-47}}

\figcaption[f46.ps]{WHI B1432-16, 1200s exposure CTIO 1.5m in R.
Approx 4 arc minutes square. \label{WHIB1432-16}}

\figcaption[f47.ps]{WHI B1444-16, 1200s exposure CTIO 1.5m in R.
Approx. 2 arc minutes square. \label{WHIB1444-16}}

\figcaption[f48.ps]{WHI B1505-67, 1200s exposure CTIO 1.5m in R.
Approx 4 arc minutes square. \label{WHIB1505-67}}

\figcaption[f49.ps]{WHI B1517-41, 1200s exposure CTIO 1.5m in R.
Approx. 8 arc minutes square.  The nebulosity is faint, general
and ill-defined.  \label{WHIB1517-41}}

\figcaption[f50.ps]{WHI B1603-04, 1200s exposure CTIO 1.5m in R.
Approx. 4 arc minutes square. \label{WHIB1603-04}}

\figcaption[f51.ps]{PGC 057387, 1200s exposure CTIO 1.5m in R.
Approx. 4 arc minutes square. \label{PGC057387}}

\figcaption[f52.ps]{WHI B1619-67, 1200s exposure CTIO 1.5m in R.
Approx. 8 arc minutes square.  \label{WHIB1619-67}}

\figcaption[f53.ps]{PGC 058179, 1200s exposure CTIO 1.5m in R.  Approx.
4 arc minutes square. \label{PGC 058179}}

\figcaption[f54.ps]{WHI B1728-08, 1800s exposure in H$\alpha$ 
INT (2.5m).  Approx 3 arc minutes square. \label{
WHIB1728-08}}

\figcaption[f55.ps]{WHI B1751-07, 600s exposure in R, INT 
(2.5m).  Approx. 3 arc minutes square.  \label{WHIB1751-07}}


\figcaption[f56.ps]{PGC 062147, 1200s exposure CTIO 1.5m in R.
Approx. 4 arc minutes square. \label{PGC062147}}

\figcaption[f57.ps]{ESO 458G011, two 1200s exposures combined,
CTIO 1.5m in R.  Approx. 4 arc minutes square. \label{ESO458G011}}

\figcaption[f58.ps]{WHI B1919-04, 600s exposure INT (2.5m) in R.
Approx. 3 arc minutes square. \label{WHIB1919-04}}

\figcaption[f59.ps]{WHI B1952-04, 660s exposure INT (2.5m) in R.
Approx. 3 arc minutes square. \label{WHIB1952-04}}

\figcaption[f60.ps]{ESO 027G002, 1200s exposure CTIO 1.5m in R.
Approx. 2 arc minutes square. \label{ESO027G002}}

\figcaption[f61.ps]{WHI B2212-10, 600s exposure INT (2.5m) in R.
Approx. 3 arc minutes square. \label{WHIB2212-10}}

\figcaption[f62.ps]{ESO 468G020, two 1200s exposures combined,
CTIO 1.5m in R.  Approx. 4 arc minutes square. \label{ESO468G020}}

\figcaption[f63.ps]{WHI B2317-32, 1200s exposure CTIO 1.5m in R.
Approx. 2 arc minutes square. \label{WHIB2317-32}}

\figcaption[f64.ps]{SC 18, 1200s exposure CTIO 1.5m in R.
Approx. 4 arc minutes square. \label{SC18}}

\figcaption[f65.ps]{ESO 293G035, 1200s exposure, CTIO 1.5m in R.
Approx. 2 arc minutes square. \label{E293}}
 
\figcaption[f66.ps]{LGS 2, 900s exposure INT (2.5m) in R.
Approx. 6 arc minutes square. \label{LGS2}}

\figcaption[f67.ps]{SC 24, 1200s exposure CTIO 1.5m in R.
Approx. 4 arc minutes square. \label{SC24}}

\figcaption[f68.ps]{ESO 352G002, 1200s exposure CTIO 1.5m in R.
Approx. 2 arc minutes square. \label{E352-2}}

\figcaption[f69.ps]{A 0107+01, 1200s exposure CTIO 1.5m in R.
Approx. 8 arc minutes square. \label{A0107+01}}

\figcaption[f70.ps]{ESO 416G012, 1200s exposure CTIO 1.5m in R.
Approx. 4 arc minutes square. \label{E416-12}}

\figcaption[f71.ps]{IC 1947, 1200s exposure CTIO 1.5m in R.  Approx.
4 arc minutes square. \label{IC1947}}


\figcaption[f72.ps]{ESO 056G019, 1800s exposure CTIO 1.5m in
H$\alpha$.  Approx. 8 arc minutes square. \label{E056G19Ha}}

\figcaption[f73.ps]{ESO 318G013, three 1200s exposures (one each
in V, R and I) combined, CTIO 1.5m.  Approx. 2 arc minutes square.
\label{E318G13W}}

\figcaption[f74.ps]{ESO 318G013, 1800s exposure in H$\alpha$
divided by 1200s exposure in R, CTIO 1.5m.  (Positive image; the
light areas are emission regions.)  Approx. 2 arc minutes square.
\label{E318G13div}}

\figcaption[f75.ps]{ISI 1106+0257, 1200s exposure CTIO 1.5m in
R.  Approx. 2 arc minutes square. \label{1106+0258}}

\figcaption[f76.ps]{ESO 269G070, 1200s exposure CTIO 1.5m in R.
Approx. 4 arc minutes square.  \label{E269G70}}

\figcaption[f77.ps]{IC 4247, three 1200s exposures (one each
in V, R and I) combined, CTIO 1.5m.  Approx. 2 arc minutes square.
\label{IC4247}}

\figcaption[f78.ps]{NGC 5237, 1200s exposure CTIO 1.5m in R.
Approx. 4 arc minutes square. \label{NGC5237}}

\figcaption[f79.ps]{IC 4739, 1200s exposure CTIO 1.5m in R.
Approx. 4 arc minutes square. \label{IC4739}}

\figcaption[f80.ps]{IC 4789, 1200s exposure CTIO 1.5m in R.
Approx. 2 arc minutes square. \label{IC4789}}

\figcaption[f81.ps]{IC 4937, 1200s exposure CTIO 1.5m in R.
Approx. 2 arc minutes square. \label{IC4937}}

\figcaption[f82.ps]{IC 5026, 1200s exposure CTIO 1.5m in R.
Approx. 4 arc minutes square. \label{IC5026}}

\figcaption[f83.ps]{A 2259+12, 1200s exposure CTIO 1.5m in R.
Approx. 1 arc minute square. \label{A2259+12}}

\figcaption[f84.ps]{ESO 347G008, 1200s exposure CTIO 1.5m in R.
Approx. 4 arc minutes square. \label{ESO347G008}}

\end{document}